\documentstyle{article}
\begin{document}
\LARGE
\begin{center}
\bf Dimensionality and the Cosmological Constant
\vspace*{0.7in}

\large \rm Zhong Chao Wu

\vspace*{0.3in}

Dept. of Physics, Zhejiang University of Technology,

Hangzhou 310032, P.R. China

(Dec. 10, 2004)

\vspace*{0.55in}
\large
\bf
Abstract
\end{center}
\vspace*{.1in} \rm \normalsize \vspace*{0.1in}

In the Kaluza-Klein model with a cosmological constant and a flux,
the external spacetime and its dimension of the created universe
from a $S^s \times S^{n-s}$ seed instanton can be identified in
quantum cosmology. One can also show that in the internal space the
effective cosmological constant is most probably zero.

\vspace*{0.6in}

PACS number(s): 04.65.+c, 11.30.Pb, 04.60.+n, 04.70.Dy


\vspace*{0.6in} e-mail: zcwu@zjut.edu.cn

\pagebreak

\rm

\normalsize

Cosmology is a branch of theoretical physics. The object of its
study is unique, it is the universe. This fact has led to the
First Cause problem, to which No-boundary universe has provided
the most satisfactory answer [1]. Within this framework, many
cosmological issues have been reexamined, e.g., the isotropy,
flatness, inflation, structure, time arrow and  primordial black
hole problems. In this article we will study the dimensionality
and the cosmological constant in a toy model.

It is assumed that the 4-dimensional spacetime of the universe we
live in (the so-called external spacetime) is obtained through a
dimensional reduction from a higher dimensional spacetime. This
scenario has been revived many times, for example, in the
frameworks of nonabelian gauge theory, extended supergravity and
braneworld.

In the Kaluza-Klein model, the $n-$dimensional spacetime is a
product of a $s-$dimensional manifold $M^s$ and a
$n-s-$dimensional manifold $M^{n-s}$. Many studies have been done
to show how to decompose $M$ into the product of an internal and
an external space in the classical framework. In general, it is
impossible to discriminate these factor spaces in this framework
unless one appeals to the Anthropic Principle [2].

We shall take the Freund-Rubin model as our toy model [3]. The
matter content of the universe is an antisymmetric tensor field
$A^{\alpha_1 \dots \alpha_{s-1}}$ of rank $s-1$, the so-called
flux. Its field strength is a completely antisymmetric tensor
$F^{\alpha_1 \dots \alpha_s}$. For the special case $s=2$ the
matter field becomes Maxwell. Recently, this model has also
attracted attention in the violation of the conjectured
``N-bound'' in quantum gravity [4]. Our motivation in this paper
is quite different.

The Lorentzian action can be written as
\begin{equation}
I_{lorentz} = \frac{1}{16\pi} \int_M \left (R - 2\Lambda
-\frac{8\pi}{s}F^2 \right ) + \frac{1}{8\pi} \int_{\partial M} K,
\end{equation}
where $\Lambda$ is the cosmological constant, $R$ is the scalar curvature of
the spacetime $M$ and $K$ is the extrinsic curvature of its boundary $\partial M$.

In the no-boundary universe [1], the wave
function of the universe is defined by the path integral over all compact
manifolds with the argument of the wave function as the only
boundary. The main contribution to the path integral comes from
the instanton, that is the stationary action solution. This is the so-called $WKB$ approximation.
The instanton should obey the Einstein equation
\begin{equation}
R^{\mu \nu} - \frac{1}{2} g^{\mu \nu} R + \Lambda g^{\mu \nu} =  8\pi  \theta^{\mu \nu},
\end{equation}
where the energy momentum tensor $\theta^{\mu \nu}$ is
\begin{equation}
\theta^{\mu \nu} = F_{\alpha_1 \dots \alpha_{s-1}}^{\;\;\;\;
\;\;\;\;\; \mu} F^{\alpha_1 \dots \alpha_{s-1} \nu} - \frac{1}{2s}
F_{\alpha_1 \dots \alpha_s}F^{\alpha_1 \dots \alpha_s}g^{\mu\nu}
\end{equation}
and the field equation for the flux
\begin{equation}
g^{-1/2}\partial_\mu (g^{1/2} F^{\mu \alpha_2\dots \alpha_s})=0.
\end{equation}

We use indices $m, \dots$ for the manifold $M^s$ and $\bar{m},
\dots$ for $M^{d-s}$, respectively. We assume that $M^s$ and
$(M^{d-s})$ are  topologically spheres, and only components of the
field $F$ with all unbarred indices can be nonzero. From de Rham
cohomology, there exists unique harmonics in $S^s$ [5], i.e, the
solution to the field equation (4)
\begin{equation}
F^{\alpha_1 \dots \alpha_s} =  \kappa \epsilon^{\alpha_1 \dots
\alpha_s}(s!g_s)^{-1/2},
\end{equation}
where $g_s$ is the determinant of the metric of $M_s$, $\kappa$ is
a charge constant. Since $\kappa$ is a continuous parameter, it is
expected that the instanton we are going to construct must be a
constrained one
[7].

For the instanton solution, the fact that dimension $s$ of the factor space
$S^s$ is the same as the rank of the given flux strength is crucial, since
otherwise  there is no nontrivial solution under our ansatz due to de
Rham cohomology [6].

From above one can derive the scalar curvature for each factor space
\begin{equation}
R_s =  \frac{(n-s-1)8\pi \kappa^2 }{n-2} + \frac{2s\Lambda}{n-2}
\end{equation}
and
\begin{equation}
R_{n-s} = - \frac{(s-1)(n-s)8\pi \kappa^2}{s(n-2)}+ \frac{2(n-s)\Lambda}{n-2}.
\end{equation}
It appears that the $F$ field behaves as an extra cosmological constant,
which is anisotropic with respect to the factor spaces. It turns out
that the metrics of the factor spacetimes should be Einstein.

At first, we assume the universe we live in to be most probable
with a given charge $\kappa$, and at the
$WKB$ level, the relative creation probability
of the universe is exponential to the negative of the Euclidean
action of the seed instanton. Since the action is proportional to the
product of the volumes of the two factor manifolds, the maximization
of the volumes can be realized only by the manifolds with maximum
symmetries.

At this moment, we assume neither of $R_s$ nor $R_{n-s}$ vanishes,
then the compact instanton metric should be the product $S^s
\times S^{n-s}$. The metric signature of $S^s (S^{n-s})$ depends
on the signature of $R_s (R_{n-s})$. A $m-$dimensional hyperboloid
$H^m$ with positive definite signature can be obtained via
analytic continuation from the metric of $S^m$ with negative
definite signature. From $S^m$ and $H^m$ with positive definite
signature, one can obtain the $m-$dimensional de Sitter space (
$dS_m$) and anti-de Sitter space ($AdS_m$) with the Lorentzian
signature $(-, +, \dots,+)$, respectively, via an analytic
continuation of an imaginary time coordinate in these metrics [6].

The external space we live in is associated with a time
coordinate. Therefore, one can obtain an universe which is a
product of the $dS_m (AdS_m)$ external space and the $S^{n-m}
(H^{n-m})$ internal space, $(m = s, n-s)$. It is noted that
$H^{n-m}$ can be compactified by a discrete isometry group.

From the same instanton, it seems that one can get either a $s-$
or a $n-s-$dimensional external spacetime. To discriminate these
two possibilities, one has to compare their creation
probabilities, since we assume that the universe we observe is
most probable.

The relative creation probability of the universe is [1]
\begin{equation}
P =\Psi^* \cdot \Psi \approx \exp (-I) ,
\end{equation}
where $\Psi$ is the wave function of the configuration at the
quantum transition surface. The configuration is the metric and
the matter field at the equator. $I$ is the Euclidean action of
the instanton. It is worth emphasizing that the seed instanton is
constructed by joining its south part of the manifold and its time
reversal, the north part of the manifold.

In the Lorentzian regime, the probability of a quantum state is
independent of the representation, one can use any representation
among the canonical conjugate variables. On the other hand, the
right representation is crucial for probability in the Euclidean
regime, especially for the creation probability of the universe
expressed by formula (8).

In the earlier research of quantum cosmology, people only studied
the creation scenario with a regular seed instanton. The
representation problem did not emerge explicitly. However, for a
more realistic cosmological model, the concept of a constrained
instanton is inevitable [7]. The criteria for the right
representation in the creation scenario is that the configuration
of the wave function should be continuous across the equator, i.e.
the quantum transition surface in the constrained instaton. Among
a pair of canonical conjugate variables, there must exist one
right variable for the representation. The wave function with the
right representation can be obtained through a canonical transform
from that with the wrong representation. Under the variable
change, the wave function is subject to a Fourier transform in the
Lorentzian regime, it is equivalent to the Legendre transform for
the action in the path integral approach. At the $WKB$ level, in
the Euclidean regime this corresponds to a Legendre transform for
the instanton action. The Legendre term at the equator will change
the probability value in Eq. (8). For a regular instanton, one
member of any pair of canonical conjugate variables must vanish at
the equator, so does the Legendre term. This is why the
representation problem did not bother in the earlier years of
study.

It is noted that
the action (1) is given under the condition that at the
boundary $\partial M$ the metric and the tensor field $A^{\alpha_1
\dots \alpha_{s-1}}$ are given. The action is invariant under the
gauge transformation
\begin{equation}
A_{\alpha_1
\dots \alpha_{s-1}} \longrightarrow A_{\alpha_1
\dots \alpha_{s-1}} + \partial_{[\alpha_1}\lambda_{\alpha_2
\dots \alpha_{s-1}]}.
\end{equation}
For our convenience, we can select a gauge such that there is only
one nonzero component  $A^{2 \dots s}$, where the index $1$
associated with the time coordinate is excluded [6]. There is no
way to find a gauge in which the field $A_{2 \dots s}$ is regular
for the whole manifold $S^s$. Therefore, the field $A$ must
subject to a discontinuity across the equator of the instanton.
For the wave function of the universe the gauge freedom has been
frozen. It means field $A$ is not a right representation. Instead,
one has to use the field strength $F$ representation.

In obtaining the Euclidean action from the Lorentzian action (1)
there exists a sign ubiquity associated with the factor
$\sqrt{-g_n}$. This can be dispelled by the following requirement.
The scalar curvature can be decomposed as $R_n = R_{n-s} + R_s$,
the $R$ term in the action associated with the external space must
be negative, such that the primordial fluctuation around the
minisuperspace background can take the minimum excitation state
[8].

If the external space is associated with the factor space
$S^{n-s}$, since the flux is living in $S^s$, there is no
discontinuity at the equator, and then the action is
\begin{equation}
I_{n-s} = \frac{\eta_{n-s} V_s V_{n-s}}{16 \pi} \left ( \frac{8\pi
\kappa^2(n-2s)}{s(n-2)} - \frac{8 \pi \kappa^2}{s} + \frac{4
    \Lambda}{n-2} \right ),
\end{equation}
where $\eta_{n-s} =-1 (+1)$, if $R_{n-s}$ is positive (negative), $V_m$ is the volume of $S^m$ with radius $L_m$
\begin{equation}
V_m = \frac{2\pi^{(m+1)/2}L_m^m}{\Gamma((m+1)/2)}.
\end{equation}

If the external space is associated with the factor space $S^s$, then
the Euclidean version of (1) is
\begin{equation}
I_{s}^\prime = \frac{\eta_{s} V_s V_{n-s}}{16 \pi} \left ( \frac{8\pi
\kappa^2(n-2s)}{s(n-2)} - \frac{8 \pi \kappa^2}{s} + \frac{4
    \Lambda}{n-2} \right ),
\end{equation}
where $\eta_s$ is defined in a similar way. Since the quantum
transition occurs at the equator of $S_s$, one has to replace
$A_{2...s}$ variable by its canonical momentum $P^{2...s}$ under
our minisuperspace ansatz, $P^{2...s}$ is defined as
\begin{equation}
P^{2 \dots s} = -\eta_s \int_\Sigma (s-1)! F^{1 \dots s},
\end{equation}
where $\Sigma$ denotes the equator.

The Legendre transform introduces an extra term into the above
Euclidean action
\begin{equation}
I_{Legendre} =- 2\eta_s A_{2...s}P^{2...s}= \frac{\eta_s  V_sV_{n-s} \kappa^2}{s},
\end{equation}
the factor 2 is due to the two sides of the equator. There is some
subtlety for the location of the equator or the quantum transition
surface in the space $S_s$ of negative definite metric signature,
but the formula (14) remains intact [6].

The total action should be
\begin{equation}
I_{s} = \frac{\eta_{s} V_s V_{n-s}}{16 \pi} \left ( \frac{8\pi
\kappa^2(n-2s)}{s(n-2)} + \frac{8 \pi \kappa^2}{s} + \frac{4
    \Lambda}{n-2} \right ).
\end{equation}

There are four possibilities:

(i) For the case $(R_s < 0, R_{n-s} <0)$, if $\kappa$ is
imaginary, then from (8)(11) and (15) one can see that creation
probability for $s-$dimensional external spacetime is
exponentially dominating over that for $n-s-$dimensional
counterpart, i.e, the apparent spacetime dimension is $s$. If
$\kappa$ is real, then the situation should be opposite. It is
noted, in the Lorentzian regime, that the field strength is real
for both cases.

(ii) For the case $(R_s > 0, R_{n-s} >0)$, if $\kappa$ is real
(imaginary), the
    external spacetime is $s-$dimensional $(n-s-$dimensional).

(iii) For the case $(R_s > 0, R_{n-s} <0)$, if $\Lambda > - 2\pi\kappa^2
    (n-2s)s^{-1}$, the dimension is $s$, otherwise, it is $n-s$.

(iv) For the case $(R_s < 0, R_{n-s}>0)$, the situation is
opposite to that  in case (iii).

When we consider the case with a zero cosmological constant. There
are only two possibilities (iii) and (iv). The dimension of the
external spacetime must be $min(s, n-s)$. That is, the dimension
of the internal space must not be lower than that of the external
space. The model with $n=11$ and $s =4$ is more realistic, it is
associated with $d=11$ supergravity, the case with a more general
flux configuration is also discussed [6].

In the above discussion the parameter $\kappa$ is fixed. Since we
suppose to live in a most probable universe, we can relax this
condition now. This corresponds to finding the instanton with
lowest Euclidean action for a given $\Lambda$. From the form of
actions (11) and (15), one can see that this may be possible as
the volume of one factor space blows up, i.e, either the scalar
curvature $R_s$ or $R_{n-s}$ vanishes.

If $R_s$ approaches zero, then we have
\begin{equation}
\kappa^2 \approx - \frac{s\Lambda}{4\pi (n-s-1)},
\end{equation}
\begin{equation}
R_{n-s} \approx \frac{2(n-s)\Lambda
}{n-s-1},
\end{equation}
\begin{equation}
I_{n-s}\approx  \frac{\eta_{n-s} V_s V_{n-s}\Lambda}{4 \pi(n-s-1)}
\longrightarrow - \infty
\end{equation}
and
\begin{equation}
I_{s}\approx O(R_s)V_sV_{n-s}.
\end{equation}
It turns out from (8)(18) and (19) that the probability for the
internal spacetime to be nearly flat and the external space to be
$dS_{n-s}$ for $ \Lambda > 0$ or $AdS_{n-s}$ for $\Lambda <0$ is
exponentially dominating.

If  $R_{n-s}$ approaches zero, then we have
\begin{equation}
\kappa^2 \approx  \frac{s\Lambda}{4\pi (s-1)},
\end{equation}
\begin{equation}
R_{s} \approx \frac{2s\Lambda}{s-1},
\end{equation}
\begin{equation}
I_{n-s}\approx  O(R_{n-s})V_sV_{n-s}
\end{equation}
and
\begin{equation}
I_{s} \approx  \frac{\eta_{s} V_s V_{n-s}\Lambda}{4 \pi(s-1)}
\longrightarrow - \infty .
\end{equation}
By the same argument, if $\Lambda > 0 $ or $(\Lambda < 0)$, then the
most probable external spacetime is $dS_s$ or $(AdS_s)$, and the
internal space is nearly flat with a very small effective
cosmological constant.

In summary, we have discussed the quantum Kaluza-Klein model with
a flux and a cosmological constant. We assume that the seed
instaton is topologically $S^s \times S^{n-s}$. The dimensionality
of the external spacetime can be identified in quantum cosmology.
We find that the most probable spacetime is of a nearly flat
internal space, that is the internal space is of a zero effective
cosmological constant, regardless of the value $\Lambda$. This
result can be used to be compared with the earlier argument that
the cosmological constant of an ordinary 4-dimensional spacetime
is most probably zero [9].

Even though from this model one can draw some concrete conclusions
about the dimension and the cosmological constant, it seems that
the model is too simple to be realistic, one has to deal with more
sophisticated models in the future.

\vspace*{0.2in}

\bf References:

\vspace*{0.1in}
\rm

1. J.B. Hartle and S.W. Hawking, \it Phys. Rev. \rm \bf D\rm
\underline{28} 2960 (1983).

2. S.W. Hawking, \it The Universe in a Nutshell  \rm (Bantam
Books, New York) chap 3 (2001).

3. G.O. Freund and M.A. Rubin,  \it Phys. Lett. \bf B\rm\underline{97}
233 (1980).

4. R. Bousso, O. DeWolfe and R.C. Myers, \it
   Found. Phys. \rm\underline{33} 297 (2003).

5. T. Eguchi, P.B. Gilkey and A.J. Hanson \it Phys. Rep.
\rm\underline{66} 213 (1980).

6. Z.C. Wu, \it Phys. Rev. \rm\bf D\rm\underline{31}, 3079 (1985).
Z.C. Wu,  \it Gene. Rel. Grav.  \rm \underline{34} 1121 (2002),
hep-th/0105021. Z.C. Wu, \it Phys. Lett. \bf B\rm \underline{585}
6 (2004), hep-th/0309178.

7.  Z.C. Wu, \it Int. J. Mod. Phys. \rm \bf D\rm \underline{6} 199
(1997), gr-qc/9801020. Z.C. Wu, \it Phys. Lett. \bf B\rm
\underline{445} 174 (1999), gr-qc/9810012. Z.C. Wu, \it Gene. Rel.
Grav.  \rm \underline{30} 1639 (1998), hep-th/9803121. Z.C. Wu,
\it Phys. Lett. \bf B\rm \underline{613} 1 (2005), gr-qc/0412041.

8. J.J. Halliwell and S.W. Hawking, \it Phys. Rev. \rm \bf
D\rm\underline{31}, 346 (1985).

9. S.W. Hawking, \it Phys. Lett. \bf B\rm
\underline{134} 403 (1984).

\end{document}